\documentclass[12pt,epsfig]{article}
\usepackage{epsfig}
\usepackage{color}

\catcode`\@=11
\def\@citex[#1]#2{%
\if@filesw \immediate \write \@auxout {\string \citation {#2}}\fi
\@tempcntb\m@ne \let\@h@ld\relax \def\@citea{}%
\@cite{%
  \@for \@citeb:=#2\do {%
    \@ifundefined {b@\@citeb}%
      {\@h@ld\@citea\@tempcntb\m@ne{\bf ?}%
      \@warning {Citation `\@citeb ' on page \thepage \space undefined}}%
      {\@tempcnta\@tempcntb \advance\@tempcnta\@ne%
      \@tempcntb\number\csname b@\@citeb \endcsname \relax%
      \ifnum\@tempcnta=\@tempcntb 
        \ifx\@h@ld\relax%
          \edef \@h@ld{\@citea\csname b@\@citeb\endcsname}%
        \else%
          \edef\@h@ld{\ifmmode{-}\else--\fi\csname b@\@citeb\endcsname}%
        \fi%
      \else
        \@h@ld\@citea\csname b@\@citeb \endcsname%
        \let\@h@ld\relax%
      \fi}%
    \def\@citea{,\penalty\@highpenalty\,}%
  }\@h@ld
}{#1}}

\def\@citeb#1#2{{[#1]\if@tempswa , #2\fi}}
\def\@citeu#1#2{{$^{#1}$\if@tempswa , #2\fi }}
\def\@citep#1#2{{#1\if@tempswa , #2\fi}}

\def\bcites{         
        \catcode`\@=11
        \let\@cite=\@citeb
        \catcode`\@=12
}

\def\upcites{         
        \catcode`\@=11
        \let\@cite=\@citeu
        \catcode`\@=12
}

\def\plaincites{      
        \catcode`\@=11
        \let\@cite=\@citep
        \catcode`\@=12
}

\def\){\right)}
\def\({\left( }
\def\]{\right] }
\def\[{\left[ }

\def\no{\nonumber \\}

\def\be{\begin{equation}}
\def\ee{\end{equation}}
\def\ba{\begin{eqnarray}}
\def\ea{\end{eqnarray}}
\def\bea{\begin{eqnarray}}
\def\eea{\end{eqnarray}}
\def\no{\nonumber \\}
\def\a{\alpha}
\def\b{\beta}

\def\n{\nu}

\begin{document}
\begin{titlepage}

\begin{center}
{\LARGE \bf Tests of Quantum Gravity\\via Generalized Uncertainty Principle
}
\vskip 1cm {\bf Yumi Ko\footnote{koyumi@khu.ac.kr}, Sunggeun Lee\footnote{
sglee@photon.khu.ac.kr} and Soonkeon Nam\footnote{
nam@khu.ac.kr}}\end{center} \centerline{\it
Department of Physics and Research Institute for Basic Sciences,}
\centerline{\it Kyung Hee University, Seoul, 130-701, Korea} \vskip
1cm


\begin{center}{\bf ABSTRACT}\end{center}
In this  paper we propose a way of determining the subleading
corrections to the Bekenstein-Hawking black hole entropy by
considering a modified generalized uncertainty principle with two
parameters. In the context of modified generalized uncertainty principle, coefficients of the
correction terms of black hole entropy are written in terms of
combination of the parameters. We also calculate corrections to the
Stefan-Boltzman law of Hawking radiation corresponding to modified
generalized uncertainty principle. By comparing the entropy with one from
black holes in string theory compactified on a Calabi-Yau manifold, we
point out that the topological information of the compactified space can not easily be related to the parameters in
modified generalized uncertainty principle. 

\renewcommand{\theequation}{\thesection.\arabic{equation}}
\end{titlepage}

\newpage

\section{Introduction}
The calculation of the black hole entropy is the first test of all
quantum gravity theories. The leading term in the black hole entropy
is the Bekenstein-Hawking (BH) entropy formula, which says that the
entropy is proportional to the horizon area $A$ of the black hole, i.e. $S_{BH}=A/4 G
$ {\cite{Bekenstein,Hawking}}. In general,
one expects that the black hole entropy is a function of the horizon
area $A$ due to holography {\cite{'tHooft,susskind}}.
One of the most promising candidate of quantum gravity is string theory, and the
BH entropy was correctly reproduced 
in terms of D-brane charges {\cite {sv,cm}}.
Recently, in ${\mathcal N}=2$ or ${\mathcal N}=4$ supergravity {\cite{mohaupt}} which
is a low energy theory of string compactified on Calabi-Yau manifold, the entropy of black holes
was calculated in terms of charges fixed by its moduli through
attractor mechanism {\cite{fks,stringentro}}. 
Following the proposal by Ooguri, Strominger and Vafa (OSV) {\cite{osv}},
for certain cases the black hole entropy was calculated exactly including a 
logarithmic term.   
The topological information of Calabi-Yau manifold is encoded
in the correction terms of the entropy.          
In this paper we investigate this problem in the context of
Generalized Uncertainty Principle (GUP). 
\be
\Delta x \ge {\hbar \over {\Delta p}} + \a L_P^2 {{\Delta p} \over
  \hbar} + \b L_P^4 {{\Delta p^3} \over \hbar^3}
\ee
One can say that GUP is a gravitational generalization of Heisenberg uncertainty
principle. It first appeared in the context of string
theory {\cite{Veneziano:1986zf,Gross:1987ar,Amati:1988tn,Konishi:1989wk}},
which has a fundamental length
scale $\lambda_s$ that gives the intrinsic limitation of resolution on probing processes
using strings. We can also obtain GUP by considering gravitational
interaction between quantum particles like electron and photon in which the
gravitational effect is no longer small enough to ignore
{\cite{adler99}}. Then the right hand side of the standard
uncertainty relation $\Delta x \ge \hbar/\Delta p$ gains a correction proportional to  $\Delta p$ which gives
minimum value of the position uncertainty. 
We can obtain the BH entropy of the black hole with calibration factor from the standard uncertainty
relation {\cite{oh}}. If we consider GUP, the area law of the entropy is
modified and it has a logarithmic correction {\cite{adler01}}. Moreover, one can obtain
corrections in all orders to the entropy by
introducing a single parameter $\a$ in GUP {\cite{1022}}, where this
coefficient is in front of the $\Delta p$~ correction term to the
uncertainty relation. 
If we assume that string theory is a reliable quantum gravity theory, general arguments of quantum gravity should give consistent results
with those in string
theory. So far the calculations in general quantum gravity theories, however, failed to fit the
coefficients of the correction terms of black hole entropy to the ones in string
theory. For GUP with a single parameter $\a$, we can  match the
coefficient of the logarithmic term fixing $\a$ but it does not
work anymore beyond the logarithmic term. For matching beyond logarithmic subleading correction terms to the
black hole entropy obtained from GUP with the ones from string theory, we should consider a
more generalized GUP with additional parameters. 
In this paper we consider a GUP with one more correction term in
position uncertainty proportional to $(\Delta
p)^3$ with parameter $\beta$ in which the gravitational effects also
get quantum correction. We will call this generalized GUP as GUP*.  We will show that $\beta$ plays the role
of fixing the coefficient of the correction term in $1/A$. From this we
anticipate that we need at least one more parameter to fix the coefficient of
the next order correction to the entropy. If one can detect
the energy density correctly, say  by experiments, one can fix $\alpha$ and $\beta$
in GUP*. We also point out that precision measurements of
hydrogen-atom spectrum gives another way of testing GUP {\cite{lay}}. Furthermore we can compare the entropies in string
theory and GUP* such that $\alpha$ and $\beta$ are related to
the topological quantities of Calabi-Yau manifold. In this way we
might be able to estimate the shape and topology of the extra dimensions experimentally and project out various string
vacua. However we will argue that direct comparison of GUP* results
with those of string theory has some difficulties. 

In Section 2, we will give a short review of GUP and see heuristically how we can
add next order corrections in GUP. In Section 3, we calculate the
black hole entropy from GUP* and see that the Stefan-Boltzman law of
Hawking radiation has
corrections in terms of $\a$ and $\b$. Then we
show the expected values of the $\a$, and $\b$ comparing with the
string result in Section 4. Finally in Section 5, we conclude with some
comments.

\section{GUP in String Theory and Quantum Gravity}

In string theory, due to the fundamental
length scale of the string $\lambda_s$, there is a restriction
in probing small distances. Hence the standard Heisenberg's uncertainty relation is modified to
\be
\Delta x \Delta p \geq \hbar + {\rm const.} ~\alpha' (\Delta p)^2,~~
\alpha'=\lambda_s^2 . \label{ups}
\ee
Note that the minimum bound of $\Delta x$, which is equal to $\lambda_s$, is encoded
in the right hand side of the relation {\cite{Veneziano:1986zf}}.
This relation can also be reproduced when we probe an electron
by a photon and also consider the gravitational effect of the
photon. Following Ref.{\cite{adler99}} it is easily shown that the position
uncertainty is influenced by the gravitational interaction between an
electron and a  photon, which is treated as a classical particle with effective
mass as follows:  
\be
\Delta x_G \approx {G \Delta p\over c^3}=L_P^2 {\Delta p\over \hbar} ,
~~~L_P =\sqrt{{G\hbar}\over c^3}.
\ee
Here $L_P$ is the Planck length.
Then the original Heisenberg uncertainty relation
$\Delta x \ge \hbar/\Delta p$ is modified to
\be
\Delta x \ge {\hbar \over \Delta p} +\a L_P^2 {\Delta p\over
  \hbar} ,
\label{gup}
\ee
where $\a$ is a dimensionless constant.
We call this relation as  Generalized Uncertainty Principle (GUP). Note
that $\Delta x$ has a minimum bound $2\sqrt{\a} L_P$.
Recall that Eq.(\ref{ups}) derived
in string theory by considering string scattering
has similar form with GUP. In this respect $\lambda_s$ is comparable to $L_P$.
Now if imagine that as one probes deep into the electron
with higher energy, GUP may be modified further by quantum
gravity effects.
To see this, we start with gravitational potential between two masses
$m_1$ and $m_2$ at distance $r$ with long distance
quantum corrections
given by\footnote{The value of the coefficient
  $k$ can be determined, for example in perturbative dynamics {\cite{dono}}.}
\be
V(r) = -{{G m_1 m_2 }\over r} + k {{G^2 \hbar~ m_1 m_2}\over{r^3
    c^3}}+\cdots {\label{cpot}}.
\ee
Heuristically,
as shown in Ref.{\cite{adler99}}, the gravitational acceleration
between the electron and the photon caused by second term of the
potential in Eq.(\ref{cpot}) will be 
\be
\ddot{\vec{r}} = {{d^2 r} \over dt^2}\hat{r} = - 3 k {{G^2 \hbar(E/c^2)}\over {r^4 c^3}}\hat{r},
\ee
where $r$ is the distance between the particles. The characteristic interacting
time inside characteristic size $L$ will be $L/c$ and the electron with velocity
\be
\Delta v \sim -3k {{G^2 \hbar E}\over {r^4 c^5}}\biggl({L\over
  c}\biggr),
\ee
will move under gravity a distance
\be
\Delta x_G \sim \Delta v \biggl({L\over
  c}\biggr)\sim  -3k {{G^2 \hbar E}\over {r^4 c^5}} {\biggl({L\over c}
\biggr)^2}.
\ee
The electron may be anywhere in the interaction region so we take 
$r\approx L$. The energy of the photon is $E=pc$ and the uncertainty
of momentum of the electron is of order of that of
the photon. If we consider the photon which has wavelength $\lambda$, of order of $r$,
we get 
\be
\Delta x \sim -3k {{G^2 (\Delta p)^3}\over {c^6 \hbar}} = - 3k {\biggl({{G^2
  \hbar^2}\over c^6 }\biggr)} {({\Delta p})^3\over \hbar^3 } = -3k L_P^4  {({\Delta p})^3\over \hbar^3 }.
\ee
\noindent
As we see above, the quantum correction to the gravitational
potential affects GUP as well. In addition, as we consider the higher order
corrections to the gravity in order of $G$, we anticipate that
there will be corrections to GUP with higher orders in $G$. It means that on dimensional ground the even power of the $\Delta p$ will not appear
in any generalization of GUP\footnote{Even power of $\Delta p$ terms in GUP  leads to corrections to
the entropy in $\sqrt{A}$ order. Nozari et al. pointed
that this type of correction terms in GUP are excluded by the fact
that the entropy does not have $\sqrt{A}$ order corrections in
string theory. {\cite{noza}}}.
From now on, we consider GUP which contains lowest order correction to the gravitational
interaction with $G^2$, i.e. we have   
\be
\Delta x \ge {1 \over \Delta p} + \alpha G {\Delta p }+ \beta G^2 {(\Delta
  p)^3 },\label{gup2}
\ee
where $\a$ and $\b$ are dimensionless
constants. Here we represent the coefficients of the modified GUP in terms of
$G$ with $c=\hbar=1$. We will call this modified GUP as GUP*. Note that $\a$ is thought to be positive from various
literatures so far but sign of $\b$ is not
determined. According to the result from the corrected gravitational
potential in Eq.(\ref{cpot}), it is
supposed to be negative. So we need to discuss the behavior of
GUP* depending on signs of $\b$. The position uncertainty versus
the momentum uncertainty of GUP* with certain values of $\a$ and $\b$ is sketched in Figure $1$. In the case of positive $\b$,
GUP*  always has a minimum value of position uncertainty 
\be
(\Delta x)_{\rm{min}} = \sqrt{2\over 27} \Biggl[~ {{-\a^2 + 12 \b + \a\sqrt{\a^2 +
    12\b}}\over{({-\a\b + \b\sqrt{\a^2 + 12\b}})^{1/2}}~}~\Biggr] L_P ,
\ee
and it is greater than the one in GUP which is $2\sqrt{\a} L_P$. If $\b$ is
negative, however, it would be inevitable that $\Delta x$ decreases as
$\Delta p$ grows. So $\Delta x$ becomes zero at a certain point. Since
we do not want such a behavior for $\Delta x$, we would like to
restrict the range of the parameter $\a$ to be
\be
\a^2 \ge 12 |\b|,~~{\rm{for}}~ \b < 0, \label{alpharange}
\ee
so that the right hand side of GUP* has two extrema at $p=p_{\pm}$,
\be
\Delta p = p_{\pm}= \Biggl[~{{\a \pm \sqrt{\a^2 -12|\b|}}\over{6
    |\b|}}~\Biggr]^{1/2} {1\over L_P}.
\ee
Then $p_-$ will give the minimum value of $\Delta x$ 
\be
(\Delta x)_{\rm{min}} = \sqrt{2\over 27} \Biggl[ {{\a^2 + 12|\b| - \a\sqrt{\a^2 -
    12|\b|}}\over{{(\a|\b| - |\b|\sqrt{\a^2 - 12|\b|}~)^{1/2}}}}~\Biggr]
L_P \label{min2},
\ee
in region where $\Delta p$ is smaller than $p_+$. Thus we will ignore
the region where $\Delta p$ is greater than $p_+$. We expect higher order
corrections to GUP to cure this problem of $\Delta x$ becoming zero. In this
case the minimum value in Eq.(\ref{min2}) is smaller than the one in GUP. 
In the next section we will see the black
hole thermodynamics from it. 
\begin{figure}[h]\label{gupgr}
\centering{\epsfig{file=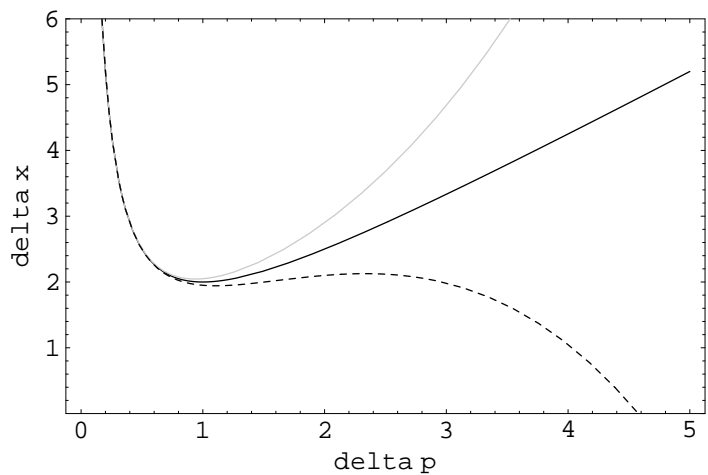}}
\caption{\footnotesize {The uncertainty in position in units of Planck length versus
  the uncertainty in momentum in units of Planck mass. The solid line
  denotes GUP with $\b=0$. Gray line and
  dashed line denotes GUP* with $\a =1$, $\b = 0.05$
  and $\a=1$, $\b=-0.05$.}}
\end{figure}
\section{GUP and Black Hole Physics}
In following discussion about the black hole entropy,
let us begin to consider a way of re-expressing GUP* in $\Delta p$ for
convenience.  
In the case of GUP in Eq.(\ref{gup}), we can obtain the exact form of $\Delta p$
expression easily as follows\footnote{We chose a negative sign to
  agree with classical result in the limit $L_P \rightarrow 0$.}
\begin{eqnarray}
\Delta p \ge { \Delta x \over { 2 \alpha G }} \biggl( ~1 - \sqrt{ ~1 - {{ 4 \alpha G } \over {(\Delta
  x)^2}} }~ \biggr)\label{dp}.
\end{eqnarray} 
In addition, regarding 
$G / (\Delta x)^2  = L_P^2 / (\Delta x)^2$ as a small
value relative to unity, we can expand the square root and obtain
\begin{eqnarray}
\Delta p \ge{ 1 \over {\Delta x}} {\biggl(~ 1 + {{ \alpha G}\over(\Delta x)^2}  + {{2 \alpha^2 G^2 }
\over(\Delta x)^4}  + \cdots \biggr)}\label{dp2}.
\end{eqnarray}
Note that this expansion form can be safely considered in
macroscopic regime {\cite{1022}}. Now, let us look at GUP*
\be
\Delta x \ge {1 \over \Delta p} + \alpha G {\Delta p }+ \beta G^2 {(\Delta
  p)^3 }.\label{gup3}
\ee 
Different from GUP in Eq.(\ref{gup}), it is not easy to be solved
analytically in $\Delta p$, so we deal with
this problem using a series solution method which was used in
obtaining Eq.(\ref{dp2}). It is reasonable to use this
method since the coefficient of the $(\Delta p)^3$ correction term is
supposed to be small. Considering $G$ as an expansion parameter, we
expect that the $\Delta p$ expression of GUP* can be written in the
vicinity of Eq.({\ref{dp2}}) as follows
\begin{equation}
\Delta p \ge p_0 + \sum^{\infty}_{n=1} G^n p_n ,
\end{equation}
\noindent
where the $p_0$ is the right hand side of Eq.(\ref{dp2}). Substituting this
to Eq.(\ref{gup3}), we read off the values of $p_n$ as
\begin{eqnarray}
p_1=0,~~p_2={\beta \over (\Delta x)^5}~,~~
p_3={{6 \alpha \beta}\over(\Delta x)^7}~,~~
p_4={{28 \alpha^2 \beta + 4 \beta^2} \over (\Delta x)^9}~,~\cdots,
\end{eqnarray} 
\noindent
and we finally have perturbatively expanded form of $\Delta p$ as
\bea
\Delta p \ge  { 1 \over {\Delta x}} \( 1 + {{\alpha G} \over {(\Delta x)^2 }}
+ {{2\alpha^2 + \beta} \over {(\Delta x)^4}}G^2 +
{{5 \alpha^3 + 6 \alpha\beta} \over {(\Delta x)^6 }}G^3 + \cdots \).\label{dp3}
\eea
It has been known that if we consider a photon as a probe of quantum
particle with position uncertainty $\Delta x$ in relativistic case, the energy of the
quautum particle have lower bound as $E \ge 1/\Delta x$ {\cite{lpb}}. Generalizing
this result to GUP* we have
\bea
E \ge  { 1 \over {\Delta x}} \( 1 + {{\alpha G} \over {(\Delta x)^2 }}
+ {{2\alpha^2 + \beta} \over {(\Delta x)^4}}G^2 + {{5 \alpha^3
+ 6 \alpha\beta} \over {(\Delta x)^6 }}G^3 + \cdots \). \label{E}
\eea
Now, let us consider the black hole entropy from GUP*. There are two
known ways of obtaining black hole entropy from uncertainty
principle. One is for a small black hole in which the black hole
evaporation takes place dominantly. In this case we translate the characteristic energy of emitted
photon regarded as the Hawking radiation from the
surface of the black hole to Hawking temperature. The
characteristic energy may be estimated from the uncertainty principle and we get black hole entropy using first thermodynamic law,
$dS= T^{-1}dE$ {\cite{oh,adler01}}. Another way is for a macroscopically large black
hole. In this case, we take an argument that the black hole
entropy is proportional to the horizon area{\cite{Bekenstein}. If we consider a
black hole absorbing a quantum particle of energy $E$ and size $R \sim
\Delta x$ in the vicinity of the horizon {\cite{chr,disper}}, the minimal increase of the horizon
area with calibration factor is given by 
\be
( \Delta A)_{\rm{min}} \ge 4({\rm{ln}}2)  L_P^2 E \Delta x , \label{amin}
\ee
 and we can
rewrite above using uncertainty principle. Finally, due to the fact that the
minimal increase of the entropy should be $\rm{ln} 2$, we can compute the
black hole entropy as we will see below. 
Let us use GUP* in the form of Eq.(\ref{dp3}) for applying above methods. 
Note that for applying GUP* for the first method, we need to know the
exact form of $\Delta p$ expression of GUP* since the Hawking 
radiation of the small black hole would be reasonably considerable at
the Planck scale. However since we obtained Eq.(\ref{dp3}) using series
solution method in the macroscopic regime, only the second method can be
used here.

Let us see the minimal increase of area from GUP*. 
Substituting Eq.(\ref{dp3}) and Eq.(\ref{E}) to Eq.(\ref{amin}), we have
\be
(\Delta A)_{min} \ge 4({\rm{ln}}2) L_P^2 \( 1 + {{\alpha G} \over {(\Delta x)^2 }}
+{{2\alpha^2 + \beta} \over {(\Delta x)^4}}G^2 + {{5 \alpha^3
+ 6 \alpha\beta} \over {(\Delta x)^6 }}G^3 + \cdots \).
\ee
Due to the fact that any particle which falls into the black hole has an intrinsic position
uncertainty of about the Schwarzschild radius $r_s = 2GM$ {\cite{adlerdas,1022}}, we choose $\Delta x
\sim 2 r_s $ and it leads to $(\Delta x)^2 \sim A / \pi$. Then we can
rewrite above as follows,
\be
( \Delta A)_{\rm{min}} \ge 4({\rm{ln}}2) {G} \[ 1 +  \alpha \({{\pi G}
\over A}\) + ( 2\alpha^2 + \beta )\({{\pi G}
\over A}\)^2 + (5\alpha^3 + 6\alpha\beta) \({{\pi G}
\over A}\)^3 + \cdots  \] .
\ee
Considering the minimal increase of entropy {\cite{chr}}, we can
relate the area with entropy as follows:
\ba
{dS \over dA} \simeq {{(\Delta S)_{\rm{min}}} \over {(\Delta A)_{\rm{min}}}}
&\simeq&{{ {{\rm{ln}}2} \over{ 4 ({\rm{ln}}2) G }{\[ 1 +  \alpha \({{\pi G}
\over A}\) + ( 2\alpha^2 + \beta )\({{\pi G}
\over A}\)^2 + \cdots  \]}}}~,
\ea
or
\be
{dS \over dA} \simeq {1\over {4 G }}{\[ 1 -  \alpha \({{\pi G}
\over A}\) - ( \alpha^2 + \beta )\({{\pi G}
\over A}\)^2 -(2\alpha^3 + 4 \alpha\beta)\({{\pi G}
\over A}\)^3+ \cdots  \]}.
\ee
Finally, by integrating above with respect to the area, we obtain the
black hole entropy with corrections to BH term:
\be
S = {A \over 4 G }-{\({{\pi\alpha} \over 4}\)}{\rm ln}\({A \over
4 G} \) + \sum^{\infty}_{n=1} c_n \({A \over
{4 G}} \)^{-n} + {\rm{const}}.~ , \label{s}
\ee
where $c_n$ are given as follows:
\begin{eqnarray*}
c_1 &=& \({{\pi} \over 4}\)^2 (\alpha^2 + \beta )~, \\
c_2 &=& \({{\pi} \over 4}\)^3 ( \alpha^3 + 2\alpha\beta)~, \\ 
c_3 &=& { \({{\pi} \over 4}\)^4}{\({ {5 \alpha^4 + 15\alpha^2 \beta + 3 \beta^2 }\over 3 }\)} \cdots~.
\end{eqnarray*}
We see that the entropy contains a logarithmic correction. In addition
using thermodyamic relation $dS = T^{-1} dM$ and taking $A=16 \pi M^2
G^2 $, we can obtain temperature $T_{\rm{GUP^*}}$ from the entropy in Eq.({\ref{s}}) in
terms of mass as follows:
\be
T_{\rm{GUP^*}} = {{M_p}^2 \over {8 \pi M }} {\( 1 +{{\a {M_P}^2 }\over{16 M^2 }}
+ {{(2\a^2 + \beta) {M_P}^4}\over{256 M^4 }} + \cdots \)}.\label{t}
\ee
Figure 2 represents the temperature of the black hole versus the black
hole mass for each case of standard uncertainty principle and GUP*.
\begin{figure}[h]\label{grmt}
\centering{\epsfig{file=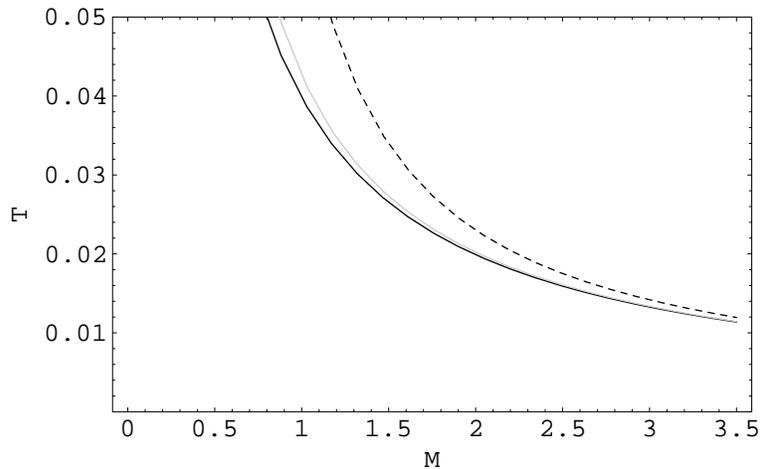}}
\caption{\footnotesize{Temperature of a black hole versus the mass in units of Planck
  mass.
 The solid line
  denotes standard uncertainty principle with $\a=0$ and $\b=0$. Gray and  dashed line denotes GUP* with $\a =1$, $\b = 0.05$
  and $\a=1$, $\b=-0.05$.
}}
\end{figure}

Now let us take a limit of $\alpha \rightarrow 0$ in  GUP*. It
is the $G^2$ correction to the standard uncertainty principle written as follows:
\be
\Delta x \ge {1 \over \Delta p}+ \beta G^2 \Delta
p^3 ~.
\ee
Eliminating $\a$ from Eqs.(\ref{s}), we have the final
form of the entropy as
\ba
S &=&  {A \over 4 G }+ \beta \({{\pi} \over 4}\)^2 \({A \over
4 G }\)^{-1 } + \beta^2  \({{\pi} \over 4}\)^4 \({A \over
4 G }\)^{-3 } + \cdots \no
&=& {A \over 4 G } + \sum_{n=1}^{\infty} {{\biggl( {{\beta
        \pi^{2}}\over {16}}\biggr)^n} \({A \over 4 G }\)^{-2n+1}}
+{\rm{const.}} 
\ea
Note that GUP with the pure $(\Delta p)^3$ ($\alpha=0$) term does not
give rise to a logarithmic term. We get only odd power terms in $1/A$. It means that
while $\b$ plays a crucial role in fixing the second correction term of the black
hole entropy, the logarithmic correction term is determined only
by $\a$. we also anticipate that we need $G^3$
corrected GUP with a $(\Delta p)^5$ term for fixing a coefficient of
third order correction term of the entropy.

Now let us consider the black hole evaporation.
Some earlier works on modification of the black hole spectrum can be
found in Refs. \cite{adler01,disper}.
As  discussed in Ref. \cite{disper}, if there is a GUP with ordinary form of
dispersion relation, one obtains a modified black body spectrum. To see
this we assume that the de Broglie wave length is modified by GUP and
takes a form as follows:
\be
\lambda \simeq {1\over p}(1+ \a G p^2 + \b G^2 p^4 ),
\ee
and it can be re-expressed in $E$ as
\be
E \simeq \n(1+ \a G \n^2 + (2 \a^2 G^2 + \b G^2 ) \n^4 ),
\ee
up to second order in $\a$ and first order in $\b$. From now on we
will follow this approximation.
Number of modes for photons in a box with edges
of length $L$ in an infinitesimal frequency interval is given by
\be
g(\nu)d\nu=8 \pi V\nu^2d\nu.
\ee
In the above, $V=L^3$ is the volume of the box. Thus,the average energy per oscillator for GUP* would be
\ba
{\bar E} &=& {E \over{e^{E\over T} -1}}\cr
&\cong&{\n\over{e^{\n\over T}}-1} \Biggl[ 1 + \a G \n^2 + \biggl\{2 \a^2 +\b -\biggl(
\a^2 (3- {\n \over {2 T}}) + \b \biggr) \biggl({{\n\over T}\over{1-
    e^{-\n\over T}}}\biggr)\biggr\} G^2 \n^4\Biggr].
\ea
Then the modified Stefan-Boltzman law takes the form 
\ba
u(T) &=& {1 \over V}\int_0^{\infty} {\bar E} g(\n) d\n \cr
&\cong&{{8 \pi^5 }\over 15}T^4 -{{320 \pi^7 \a G }\over 63} T^6 - {{64
    \pi^9(22 \a^2 +7 \b) G^2 }\over 15}T^8 - 1451520 \pi \a^2 G^2
\zeta(9) T^8 , {\label{ut}}
\ea
where $u(T)$ is the energy density and the numerical value of the zeta
function is approximately $\zeta(9) \cong 1.00201$. 
The radiation emitted by a black hole can be considered by energy
conservation equation
\be
{dM \over dt} = - A u(T_{\rm{GUP^*}}). 
\ee

Inserting Eq.(\ref{ut}) and introducing dimensionless mass $m=M/M_P$ one can rewrite it in
Planck mass unit as
\bea
{dm \over dt} \cong - {1\over { t_{ch}
     m^2 }}\biggl[ 1 + {{17 \a} \over {168 m^2}}
+\biggl\{{\b \over 512} - \a^2 \biggl({79 \over 1792} + {{42525 \zeta(9)} \over
 { 64 \pi^8}}\biggr)\biggr\} {1 \over m^4} ~ \biggr],
\eea
where the characteristic time is  $t_{ch} = {{480
  }\over{\pi^2}}T_P$, where $T_P$ is the Planck time $1/{M_P}$. From this we
  obtain a relation between time and black hole mass
\be
{t \over t_{ch}} \cong \biggl[{m^3 \over 3} + {{17 m \a}\over 168} + {{43
    \a\b}\over {832 m^3}} + \biggl\{-{\b \over 512} + \a^2 \biggl({3821 \over
  112896} + {{ 42525 \zeta(9)}\over{64 \pi^8}}\biggr)\biggr\}{1 \over
  m}\biggr]_{M/M_P}^{M_i /M_P} ,
\ee
where $M_i$ is the initial mass of the black hole. The first term of the right hand side
represents a result from the standard uncertainty relation that the
black hole evaporates to zero mass in finite time. For our case,
however, it will take infinite time for the black hole to evaporate totally. So we expect that the GUP* may prevent the black hole from
vanishing by evaporation.

\section{Black Holes in String Theory vs. GUP}
Recently, the entropy of BPS black holes
associated with D-branes wrapped on Calabi-Yau manifold in type II
string theory was computed in terms of charges fixed by its moduli at near
horizon {\cite{stringentro}}. In this case, the entropy has
correction terms including a logarithmic term. 
According to the OSV conjecture {\cite{osv}}, the number of BPS states of the black
hole is obtained by topological string partition function given in
moduli space of the Calai-Yau manifold. Therefore, the topological
informations of the extra dimension is encoded in the black hole entropy.         
Now we consider an example given by Dabholkar et
al.\cite{stringentro}. They tested OSV poposal and obtained
results that the entropy computed in each macroscopic and microscopic point
of view were matched. The resulting entropy is written
as\footnote{Here, $(p^I, q_I)$ are magnetic and electric charges
  respectively defined on the Calabi-Yau
  manifold where $I=0,1,\cdots,h^{1,1}$.  $\nu={1\over
    2}(h^{1,1}+2)$, $\hat{q}_0 = q_0 - 1/2 q_a C^{ab}(p) q_b$, $\hat{C}(p)=C(p)+c_{2a}p^a$. And
$C_{ab}(p)=C_{abc}p^c,C(p)=C_{abc}p^ap^bp^c,$
where $C_{abc}$ are the intersection numbers and $c_{2a}$ are the
components of the second Chern class as topological quantities where
$a,b,c=1,\cdots,h^{1,1}$ \cite{stringentro}.}
\bea
S&=&2\pi \sqrt{ {\hat{C}|\hat{q}_0|\over 6}} -\biggl(\nu+{1\over 2}\biggr)
\ln \Biggl(2\pi \sqrt{ {\hat{C}|\hat{q}_0|\over 6}}~\Biggr)-{{4\nu^2-1}\over 8}
\Biggl({2\pi \sqrt{ {\hat{C}|\hat{q}_0|\over 6}}}~\Biggr)^{-1} \cdots ,
\label{s1}
\eea
Thus the topological information of the
internal manifold is encoded in the black hole entropy. Identifying the first term as
$A/4G$, we have the familar form of the entropy
with a Bekenstein Hawking leading term. Let us compare this with Eq.(\ref{s}). In the sense that any quantum gravity theory should give
the same results in low energy limit, the entropy from supergravity should agree with
one from GUP*. Observing emitted radiation from black hole we can
determine $\a$ and $\b$ and they would give $\n$ by relating to
the result in string theory. We emphasize that it provides a way of knowing
information of Calabi-Yau manifold  by experiment in principle. From the
above example, the values of $\a$ and $\b$ are     
\be
\alpha={4 \over \pi }\biggl(\nu+{1\over 2}\biggr), ~~\beta=\biggl({4\over \pi}\biggr)^2
\biggl[{{1-4\n^2}\over 8}-\biggl({\n+ {1\over 2}}\biggr)^2 \biggr].
\ee
Note that since  $\n$ is positive, $\b$ is always
negative. Furthermore we always have $\a^2 < 12|\b|$. 
This violates the bound we have found in Eq.(\ref{alpharange}). 
Since our argument of GUP* was a quite general one up to order of
$G^2$, the string theory results suggest that we should include even
higher order terms.
It would be quite interesting to check whether this is really true.

\section{Summary}
In this work we have considered a generalized uncertainty principle with
higher gravitational corrections obtained by expansion in terms of the Newton's
constant $G$. Expanding position uncertainty $\Delta x$ to order of $G$
leads to the fact that the correction terms
should have only odd powers of $\Delta p$. Starting out with $G^2$ corrected
GUP as given in Eq.({\ref{gup2}}) with dimensionless parameters $\a$ and $\b$, we
obtained quantum corrected black hole entropy. It turns out that the
coefficients of correction terms of black hole entropy are given by combinations
of the parameters in GUP*. Moreover with these two parameters, we can determine
exactly up to second order correction terms of the entropy. GUP*
also leads modification of Stefan-Boltzman law for the Hawking radiation. Therefore if we can detect the radiation rate of the black
hole precisely, we can fix the values of the parameters $\a$ and $\b$.
By comparing our
black hole entropy with the one in
string theory obtained by OSV proposal, we can relate the parameters to
the topological quantities of the Calabi-Yau manifold. 
However the parameters do not satisfy constraint of GUP*, for the
example at hand. If we take this discrepancy seriously, it means that
GUP* does not easily provide a procedure to test the string theory experimentally,
and further work is necessary to demonstrate the extra dimensions and
topological properties of them experimentally. To cure this problem we
might need to generalized GUP* even further. Further
possible modification to GUP* as in Eq.(\ref{gup}) is to add some terms 
like $(\Delta p)^n$ or $(\Delta x)^n$ or some products of them.
By comparing known entropy by various theories to the entropy
from our trial uncertainty relation, one can test the quantum gravity
in remarkably simple way. 
Lastly, one challenging problem is what Heisenberg algebra gives this
kind of GUP. For GUP, corresponding Heisenberg algebra has
been known as $\kappa$-deformed type which leads to noncommutative space
time {\cite{mag,nam}} but for GUP*, the mathematical structure of
the algebra would be different.

\section{Acknowledgements}
This work of S. Nam was supported by grant
No. R01-2003-000-10391-0 from the Basic Research Program of the Korea
Science and Engineering Foundation and Center for Quantum Spacetime
(CQUeST) of Sogang University with grant number
R11-2005-021(KOSEF). This work of S. Lee was supported by grant
No. R01-2003-000-10391-0 from the Basic Research Program of the Korea
Science.

\end{document}